\documentstyle[12pt]{article}

\begin{document}
\begin{center}
{\large\bf
Generating the texture of symmetric fermion mass matrices
and anomalous hierarchy patterns for the neutrinos from
an extra abelian symmetry.}
\\
\vspace{.2in}
Elena Papageorgiu,\\
LPTHE, Universit\'e de Paris XI, B\^at. 211, F-91405 Orsay.
\end{center}
\vspace{.5in}

\begin{abstract}
\noindent
Assuming that a horizontal abelian (gauge) symmetry is
at the origin of texture zeros in the fermion mass matrices
we show how realistic mass patterns can be generated stepwise with
a small number of effective operators from GUT's and string theory.
It is interesting to note that, in the minimal scenario,
if the up quark and the down quark mass matrices are generated
simultaneously their texture is formaquivalent and contains only
two zeros. Textures with more zeros (and thus more predictability)
emerge if the up and down sectors are generated stepwise.
This last possibility leads to rather anomalous neutrino patterns
with mass degenerate neutrinos and large
$\nu_{\mu} - \nu_{\tau}$ or $\nu_e - \nu_{\tau}$ mixing
even in the absence of hierarchy in $M_R$, the
mass matrix of the righthanded neutrinos, and as required by the LSND
results if they were to be confirmed by future experiments.
In contrast if the quark mass matrices have two zero entries (or less)
the neutrino mass-mixing spectrum is independent of the
texture structure of $M_R$ -if the latter does not contain
any hierarchy of scales- and obeys the quadratic seesaw.
\end{abstract}

\vskip 2 truecm
\noindent {\bf LPTHE Orsay Preprint 95-02}
\newpage

\section{Introduction.}
In the hope to find
the underlying symmetry principle for the long list of mass and
mixing parameters which characterize the fermion spectrum and
enter in the Standard Model (SM) as free parameters there have
been many attempts since the late sixties to describe it
with a minimum of observable quantities which can be
related to a more fundamental theory [1]. It is for example known
that the observed mass and mixing hierarchies of the fermion
spectrum can be successfully described in terms of the
{\it Wolfenstein} parameter $\lambda\simeq 0.22$ which, to a good
approximation, gives the Cabbibo mixing $|V_{us}|$ [2].
Taking into account the present experimental uncertainties [3]
one finds the following mass ratios for
the up and down quarks:
\begin{equation}
m_u : m_c : m_{t_{/_{M_X}}} = \left (\xi_1(M_X)\lambda^8 :
\xi_2(M_X)\lambda^4 : 1\right ) \times m_t(M_X)
\end{equation}
\begin{equation}
m_d : m_s : m_{b_{/_{M_X}}} = \left (\xi_1^{\prime}(M_X)\lambda^4
: \xi_2^{\prime}(M_X) \lambda^2 : 1\right ) \times m_b (M_X)\,,
\end{equation}
whith $\xi_{1,2},\xi_{1,2}^{\prime}\leq {\cal O}(\lambda)$.
The scale dependent constants $\xi_1,\xi_2$ and
$\xi_1^{\prime},\xi_2^{\prime}$ contain the radiative
corrections from the running of the Yukawa couplings (since
the gauge couplings are ``family blind'') from the electroweak
scale $M_Z$ to a scale $M_X$ which can be as high as the
grand unification scale $M_G = 10^{16}$ GeV
or even the Planck scale $M_P = 10^{19}$ GeV . For the SM as
for the two-Higgs doublet models these corrections are still
within the margin of the experimental uncertainties and are to leading
order equal for the two ratios $m_u/m_t$ and $m_c/m_t$
(and for $m_d/m_b$ and $m_s/m_b$)
due to the large mass of the top quark, or more generally speaking
due to the predominance of the third-generation Yukawa couplings [4].
The same applies also to the mixing elements of the CKM matrix,
written in the Wolfenstein parametrisation as:
\begin{equation}
V_{{CKM}_{/_{M_X}}}  = \left(
\begin{array}{ccc}
1 - \lambda^2/2 & \lambda & A(M_X) \lambda^3 (\rho + i \eta) \\
 - \lambda & 1 - \lambda^2/2 & A(M_X) \lambda^2 \\
A(M_X) \lambda^3 (1 - \rho + i \eta) & - A(M_X) \lambda^2 & 1
\end{array}
\right) \,,
\end{equation}
where $A$ is of order one. The running of
the Cabbibo angle is negligible, while
the running of the small mixing elements
is incoorporated in $A(M_X)$ and is practically the same for
$|V_{13}|$ and $|V_{23}|$. In the
SM $A(M_X)$ is increasing with the energy whereas in the minimal
supersymmetric standard model (MSSM) and in the general
two-Higgs doublet case it is decreasing [4] while remaining within
the uncertainties of the present day experimental data.
Because of this it is not necessary to work within a particular
model to understand how the different powers of lambda arise,
nor to fix the scale at which the fermion mass and mixing patterns are
generated.
We shall therefore set up a framework which could be implemented
in different models, though, it may appear more natural from within the
MSSM with unification of gauge couplings and partial unification
of Yukawa couplings at $M_G$. In the latter the regularity of the
spectrum in terms of $\lambda$ is improved at the
ultraviolet, where one also notices that ratios of up quark masses and down
quark masses are related through a $\lambda^2 \leftrightarrow \lambda$
transformation. Compared to the almost equal spacing between neighbouring
quark mass levels the hierarchy in the charged
lepton sector is rather anomalous:
\begin{equation}
m_e : m_{\mu} : m_{\tau_{/_{M_X}}} = \left (\eta_1(M_X)\lambda^5 :
\eta_2(M_X)\lambda : 1\right )\times m_{\tau}(M_X) \,,
\end{equation}
where again the constants $\eta_1,\eta_2$ contain the radiative corrections.
Therefore, in models with  unification of the Yukawa couplings of the
third generation there is
no such unification for the first two generations.
Instead the following approximate relations:
\begin{equation}
m_{\tau}\simeq m_b \qquad\qquad m_e \cdot m_{\mu}\simeq
m_d \cdot m_s  \quad .
\end{equation}
hold at the unification scale for various supersymmetric grand unified
theories (SUSY GUT's) [5].

Given the fact that at low energies one has only 13 observables (six quark and
three lepton masses, the three mixing angles and the CP violating phase
of the CKM matrix) one cannot fix the entries of
the quark and lepton mass matrices $M_u$, $M_d$ and $M_e$ at $M_X$,
even by assuming that the latter are hermitian.
This has led to different {\it Ans\"atze} [1,6] in which some
of the entries are zero while the others are given in powers of
$\lambda$.
\footnote{Zeros are in fact small entries which can be neglected to leading
order.} In the quark sector the maximum number of zeros that one can
have is five (counting together those in $M_u$ and $M_d$, but without
counting symmetric entries twice) and
there are five different pairs of $M_u - M_d$ textures
\footnote{A texture in this context exhibits the power behaviour of
fermion mass matrices in terms of some scale, and, is normalized with respect
to the largest entry.} at $M_G$
which lead to masses and mixings which are compatible with the
present-day experimental values [6]. In fact,
the zeros in the mass matrices
can be thought off as ``relics'' of a new symmetry which is not
``family-blind'', while the small non-zero entries could well be
correction terms generated after symmetry breaking.
This ``old'' idea [7] has  been recently reinvestigated [8-10]
in the light of a new way of obtaining
also the successful $sin^2 \theta_W = 3/8$ result
of the canonical gauge coupling unification
which consists in extending the gauge group of the
standard model by a horizontal $U(1)_X$ factor whose
anomalies can be cancelled by the Green-Schwarz mechanism [11].

Previous investigations have shown that if the effective theory
contains higher dimension operators with a few extra scalar fields
one can generate symmetric textures with two zeros [9,10]
or asymmetric textures [12] which can successfully reproduce the
${\cal O}(\lambda^n)$ behaviour of realistic quark mass matrices.
Here we shall show that within this minimal scheme one can also generate
symmetric textures with three zeros which lead to
more predictive {\it Ansatz} for the quark mass matrices and
discuss the consequences for the neutrino spectrum in the
two cases.

\section{Generating the textures of the quark mass matrices.}
\subsection{Generating symmetric textures with two zeros.}
Let us assume the existence of a family-dependent $U(1)_X$
symmetry at $M_P$, with respect to
which the quarks and leptons carry charges $\alpha_i$ and $a_i$
respectively, where $i=1,2,3$ is the generation index.
We first consider the up quark mass matrix.
Given the predominant role played by the top quark we start with a rank-one
matrix and make a choice for the charges such
that only the (3,3) renormalizable coupling $t^c t h_1$ is allowed.
This fixes the charge of the light Higgs $h_1$ to $-2\alpha$
($\alpha\equiv \alpha_3$). We expect the other entries to be generated
by higher-dimension operators which may occur at
the string compactification level  and
contain combinations of scalar fields some of which acquire
vacuum expectation values (vev's) leading to spontaneous symmetry
breaking and large masses for the non-observable part of the spectrum. In the
most minimal scenario
one will have just one singlet field or a pair of such fields $\sigma_{\pm}$
developping equal (vev's) along a ``D-flat'' direction and carrying
opposite charges $\pm 1$. They can give rise to higher-order couplings
$q^c_{i} h_1 ({<\sigma>\over M})^{|2\alpha - \alpha_i - \alpha_j|} q_{j}$.
Notice that when the exponent is positive (negative) only the field
$\sigma_+$ $(\sigma_-)$ can contribute. The new scale
${\cal E} = {<\sigma>\over M}$ which enters in the quark mass matrix is
the ratio of the symmetry breaking scale to the scale which governs these
higher-dimension operators, and could be the string unification scale
$M_S \simeq 10^{18}$ GeV or $M_P$ or in fact any intermediate scale. The power
with which
this scale appears in the different Yukawa entries is such as to compensate the
charge of $q^c_{i} h_1 q_{j}$.
If ${\cal E}$ is a small
number one finds two universal hierarchy patterns in the texture which is
generated:
\begin{equation}\label{Q0}
M_x \sim \left(
\begin{array}{ccc}
{\cal E}^{2 |x_1|} & {\cal E}^{|x_1 + x_2|} & {\cal E}^{|x_1|}\\
{\cal E}^{|x_1 + x_2|} & {\cal E}^{2 |x_2|} & {\cal E}^{|x_2|}\\
{\cal E}^{|x_1|} & {\cal E}^{|x_2|} & 1\\
\end{array}
\right) \qquad  |x_{1,2}| = |\alpha - \alpha_{1,2}| \,,
\end{equation}
namely $m_{11} \sim m_{13}^2$ and $m_{22} \sim m_{23}^2$, where by $m_{ij}$ we
denote the
value of the entry ${ij}$.
The choices $|x_2|=1$ and $|x_1|=4$ or $|x_2|=1$ and $|x_1|=2$
lead to:
\begin{equation}\label{xx}
M_x^{(1)} \sim \left(
\begin{array}{ccc}
{\cal E}^8 & {\cal E}^3 & {\cal E}^4\\
{\cal E}^3 & {\cal E}^2 & {\cal E}\\
{\cal E}^4 & {\cal E} & 1\\
\end{array}
\right) \quad {\rm or} \quad
M_x^{(2)} \sim \left(
\begin{array}{ccc}
{\cal E}^4 & {\cal E}^3 & {\cal E}^2\\
{\cal E}^3 & {\cal E}^2 & {\cal E}\\
{\cal E}^2 & {\cal E} & 1\\
\end{array}
\right) \,.
\end{equation}
If ${\cal E}$ is of order $\lambda^2$ the two textures above
correspond to the phenomenologically acceptable {\it Ans\"atze
\`a la Fritzsch} \footnote{The original {\it Ansatz} had a zero in
the (2,2) entry.}
or {\it \`a la Giudice} for the up quark mass matrix. Notice that for
generating $M_u^{(2)}$ only
one singlet is needed while for generating  $M_u^{(1)}$ a pair of singlets with
opposite
charges are needed.
Approximating the $(1,1)$ entry in the textures of eq.(7) by a zero leads to
the well known
relation: $V_{us} \simeq \sqrt{m_d/m_s}$ [1]. Setting the $(1,3)$ entry in
$M_x^{(1)}$
or the $(1,2)$ entry in $M_x^{(2)}$ to zero leads to further mass mixing
relations [1,6].
The generation of other phenomenologically acceptable textures having a zero
also in the (2,2)
or the (2,3) entry (but not in both entries simultaneously)
necessitates a more complicated mechanism involving
extra singlets and mixing with heavy Higgses,
a case which will be discussed in the next section.

Given the up-quark textures of eq.(7) we shall try to construct
realistic textures for the down quark sector.
The assumption of symmetric mass
matrices and the $SU(2)_L$ symmetry require the equality
between the charges of the up and down quarks.
Assuming again that only the (3,3) renormalizable coupling is allowed
leads to the other light Higgs $h_2$ carrying the same charge as $h_1$.
This means that this $U(1)_X$ is anomalous and needs a cancellation mechanism.
Notice that the choice of a particular texture for $M_u$ has already
fixed the texture of $M_d$ in terms of some new scale
${\cal E}^{\prime}$ which has to be of order $\lambda$ to give the
correct mass spectrum of eq.(2). The origin of this difference in scale
${\cal E}^{\prime}\sim {\cal E}^{1/2}$ is yet unknown.
Since the {\it Ansatz} $M_u^{(2)},M_d^{(2)}$ is phenomenologically not
acceptable, the only  sucessfull
{\it Ansatz} which can be generated within this
minimal scenario for the up and down quark mass matrices is the one which
resembles the
{\it Fritzsch Ansatz} and contains
four zeros in total:
\begin{equation}\label{x1}
M_u^{i} \sim \left(
\begin{array}{ccc}
0 & \lambda^6 & 0\\
\lambda^6 & \lambda^4 & \lambda^2\\
0 & \lambda^2 & 1\\
\end{array}
\right) \quad \quad
M_d^{i} \sim \left(
\begin{array}{ccc}
0 &  \lambda^3 & 0\\
\lambda^3 &  \lambda^2 & \lambda\\
0 & \lambda & 1\\
\end{array}
\right) \,,
\end{equation}
 where we have approximated by zeros the suppressed entries.

\subsection{Generating symmetric textures with three zeros.}
In order to generate three-zero textures
additional scalar fields are needed.
In SUSY GUT's as well as in many string compactification schemes these are
commonplace.
There are Higgs multiplets $H_i$ which are responsible for the breaking of the
GUT symmetry as well as heavy singlets ${\tilde \sigma}_i$ [13] and thus higher
dimensional couplings
$q^c_{i} H ({<{\tilde \sigma}>\over M})^{|2\beta - \alpha_i - \alpha_j|}
q_{j}$,
where by $-2\beta$ we denote the charge of $H$,
which give rise to the following texture:
\begin{equation}\label{Yz}
M_z \sim \left(
\begin{array}{ccc}
{\tilde {\cal E}}^{2 |z_1|} & {\tilde {\cal E}}^{|z_1 + z_2|}
& {\tilde {\cal E}}^{|z_1 + z|}\\
{\tilde {\cal E}}^{|z_1 + z_2|} & {\tilde {\cal E}}^{2 |z_2|}
& {\tilde {\cal E}}^{|z_2 + z|}\\
{\tilde {\cal E}}^{|z_1 + z|} & {\tilde {\cal E}}^{|z_2 + z|}
& 1 + {\tilde {\cal E}}^{2 |z|}\\
\end{array}
\right) \,,
\end{equation}
whith $|z_i| = |\beta - \alpha_i|$, and
$|z| = |\beta - \alpha|$.
When the difference between the light- and heavy-Higgs charges is larger
than between the charges of the heavy Higgs and the quarks, {\it i.e.} $z\gg
z_{1,2}$, this
automatically gives suppressed (1,3) and (2,3) mass entries.
For the particular choice $z_2 = 1$ and $z_1 = 2$ (or $z_2 = 1$ and $z_1 = 4$)
\footnote{This choice requires a pair of singlets with opposite charges.}
one obtains to leading order the following texture:
\begin{equation}
M_z^{(1)} \sim \left(
\begin{array}{ccc}
0 & {\tilde {\cal E}}^3 &  0\\
{\tilde {\cal E}}^3 & {\tilde {\cal E}}^2 & 0\\
0 & 0 & 1\\
\end{array}
\right) \,.
\end{equation}
Another interesting texture structure arises when $| z_2 + z | = 1$
and $| z_1 + z_2 | = 3$, while $| z|, | z_{1,2}| \gg 0$:
\begin{equation}
M_z^{(2)} \sim \left(
\begin{array}{ccc}
0 & {\tilde {\cal E}}^3 &  0\\
{\tilde {\cal E}}^3 & 0 & {\tilde {\cal E}}\\
0 & {\tilde {\cal E}} & 1\\
\end{array}
\right) \,.
\end{equation}
The texture of eq.(11) is the original {\it Ansatz} of Fritzsch having an extra
zero in the (2,2)
entry, while the texture of eq.(10) has an extra zero in the (2,3) entry, as
compared to the two-zero texture $M_x^{(1)}$ of eq.(7).
As there are no solutions with six zeros (counting the zeros of the up quark
and down quark sector together) the most predictive {\it Ansatz} require a
mixed scenario which will
first generate $M_z^{(1)}$ or $M_z^{(2)}$  and subsequently generate the
missing entries needed for obtaining $M_x^{(1)}$ or $M_x^{(2)}$. This means
that
either the up quark and down quark sectors are
generated independently of each other or progressively one from another.
There are two possible scenarios where this can be accomplished.

\subsection{The mixed cases}
In the first scenario the up-quark mass texture
is generated first through mixing of a set of singlet fields with heavy Higgs
fields:
$u^c_{i} H ({<{\tilde \sigma}_{\pm}>\over M})^{|2\beta - \alpha_i - \alpha_j|}
u_{j}$
with ${\tilde {\cal E}} = {<{\tilde \sigma}>\over M} \sim \lambda^2$.
The down-quark mass texture is generated
when another set of singlet fields mixes with
the light Higgses:
$d^c_{i} h_1 ({<\sigma_{\pm}>\over M})^{|2\alpha - \alpha_i - \alpha_j|} d_{j}$
with ${\cal E} = {<\sigma>\over M} \sim \lambda$.
In this way one obtains the following two
set of phenomenologically acceptable solutions:
\begin{equation}
M_u^I \sim \left(
\begin{array}{ccc}
0 & \lambda^6 &  0\\
\lambda^6 & \lambda^4  & 0\\
0 & 0 & 1\\
\end{array}
\right) \quad \quad
M_d^I \sim \left(
\begin{array}{ccc}
0 & \lambda^3 & 0 \\
\lambda^3 & \lambda^2 & \lambda\\
0 & \lambda & 1\\
\end{array}
\right) \end{equation}
and
\begin{equation}\label{xx}
M_u^{II} \sim \left(
\begin{array}{ccc}
0 & \lambda^6 &  0\\
\lambda^6 & 0 & \lambda^2\\
0 & \lambda^2 & 1\\
\end{array}
\right)
\quad  \quad
M_d^{II} \sim \left(
\begin{array}{ccc}
0 & \lambda^3 & 0 \\
\lambda^3 & \lambda^2 & \lambda\\
0 & \lambda & 1\\
\end{array}
\right) \,.
\end{equation}

In the other scenario the up- and down-quark sectors
are generated independently of each other from the
couplings:
$u^c_{i} h_1 ({<\sigma_{\pm}>\over M})^{|2\alpha - \alpha_i - \alpha_j|} u_{j}$
and
$d^c_{i} H ({<{\tilde \sigma}_{\pm}>\over M})^{|2\beta - \alpha_i - \alpha_j|}
d_{j}$
respectively,
giving rise to the following two solutions:
\begin{equation}
M_u^{IV} \sim \left(
\begin{array}{ccc}
0 & \lambda^6 &  0\\
\lambda^6 & \lambda^4 & \lambda^2 \\
0 & \lambda^2 & 1\\
\end{array}
\right) \quad \quad
M_d^{IV} \sim \left(
\begin{array}{ccc}
0 & \lambda^3 & 0\\
\lambda^3 &  \lambda^2 & 0\\
0 & 0 & 1\\
\end{array}
\right) \,,
\end{equation}
and
\begin{equation}
M_u^V \sim \left(
\begin{array}{ccc}
0 & 0 & \lambda^4 \\
0 & \lambda^4 & \lambda^2 \\
\lambda^4 & \lambda^2 & 1\\
\end{array}
\right) \quad \quad
M_d^V \sim \left(
\begin{array}{ccc}
0 & \lambda^3 & 0\\
\lambda^3 &  \lambda^2 & 0\\
0 & 0 & 1\\
\end{array}
\right) \,.
\end{equation}
One should notice that the set of textures ($I,II,IV,V$)
which have emerged from  a particular choice
of the charges $\alpha_1,\alpha_2,\alpha$ and $\beta$
correspond to the approximate 5-zero texture solutions of ref.[6].
Notice however that the parametrization of the entries
of the down quark mass matrix in terms of powers of lambda
is somewhat ambiguious because in the MSSM
the factors in front are of ${\cal O}(\lambda)$ or of ${\cal
O}(\lambda^{1/2})$.
In ref.[...] it was shown that alternative textures which are {\it apriori}
possible due to this ambiguity cannot be generated within this minimal
scenario.

\section{Generating the texture of the lepton mass matrices}
\subsection{The charged lepton mass texture}
Assuming simply the gauge symmetries of the SM the $U(1)_X$ charges
$a_i$ of the
leptons  are not related to those of the quarks.
Allowing however the coupling $\tau^c \tau h_2$ leads to $a_3 = \alpha$.
Another constraint comes from the second mass relation
of eq.(5) which implies that $a_1 + a_2 = \alpha_1 + \alpha_2$.
The simplest way to satisfy this relation is to have $a_1 = \alpha_1$ and
$a_2 = \alpha_2$.
Since the early days of grand unification it is known
that in order to obtain also for the first two generations
acceptable mass relations between the charged leptons and down quarks
the (2,2) entry of $M_d$ should be multiplied by a factor $\kappa = -3$,
the other entries of $M_e$ and $M_d$ been equal [14].
In this way, though there can be no explanation of the factor minus three
in this approach
\begin{equation} M_e = M_d \qquad
M_e^{(i,I,II)} \sim \left(
\begin{array}{ccc}
0 & \lambda^3 & 0\\
\lambda^3 & \kappa\lambda^2 & \lambda\\
0 & \lambda & 1\\
\end{array}
\right)  \qquad
M_e^{(IV,V)} \sim \left(
\begin{array}{ccc}
0 & \lambda^3 & 0\\
\lambda^3 & \kappa\lambda^2 & 0\\
0 & 0 & 1\\
\end{array}
\right) \,.
\end{equation}

The alternative is a texture:
\begin{equation}
M_e^{\star}\sim \left(
\begin{array}{ccc}
0 & \lambda^3 & 0\\
\lambda^3 & \lambda & 0\\
0 & 0 & 1\\
\end{array}
\right)
\end{equation}
which can be generated from the texture $M_z$ in eq.(9) replacing
$|z_i| \to |\beta - a_i|$ and setting
$|z_2| = 1/2$ and $|z_1| = 5/2$ or $7/2$ when $|z|\gg |z_{1,2}|$.
This choice is compatible with the 5-zero solutions of eqs.(12-15)
and obviously with the 4-zero {\it Ansatz} of eq.(8) but at the expense of
introducing an extra scale in addition to the two scales needed for generating
the quark textures.

\subsection{The Dirac neutrino mass texture} \noindent
We turn now to the neutrino sector.
Again as a consequence of the $SU(2)_L$ symmetry and our symmetric
{\it Ansatz} the lefthanded and righthanded neutrinos, $\nu_i$ and
$N_i$, become charged under the $U(1)_X$ with the same charges $a_i$
as the charged leptons. Obviously the presence
of the $N_i$'s implies a larger symmetry than the minimal extension
of the SM by an extra $U(1)$ factor,
but also the assumption of symmetric mass matrices can find its
justification only in the context of a left-right symmetric
theory.
Let us first discuss the generation mechanism for Dirac neutrino mass
terms: $M^D N_i^c \nu_j$. Since these are of the same type as the
mass terms in the quark and charged lepton sector it is natural
to adopt the same approach. Then, because the charges $a_i$ have been
fixed through the charged lepton {\it Ansatz},
\begin{equation}
M^D = M_e \,.
\end{equation}
Furthermore for the choice $a_1 = \alpha_1$ and $a_2 = \alpha_2$,
which led to eq.(16), one obtains the well known GUT relations:
\begin{equation}
M^D = M_u \qquad {\rm or} \qquad M^D = M_d \,.
\end{equation}
The left equation implies four distinct textures for $M^D$:
\begin{equation}
M_u^{D(I)} \sim \left(
\begin{array}{ccc}
0 & \lambda^6 & 0\\
\lambda^6 & \lambda^4 & 0\\
0 & 0 & 1\\
\end{array}
\right)  \qquad
M_u^{D(II)} \sim \left(
\begin{array}{ccc}
0 & \lambda^6 & 0\\
\lambda^6 & 0 & \lambda^2 \\
0 & \lambda^2 & 1\\
\end{array}
\right)
\end{equation}
\begin{equation}
M_u^{D(i,IV)} \sim \left(
\begin{array}{ccc}
0 & \lambda^6 & 0\\
\lambda^6 & \lambda^4 & \lambda^2 \\
0 & \lambda^2 & 1\\
\end{array}
\right)  \qquad
M_u^{D(V)} \sim \left(
\begin{array}{ccc}
0 & 0 & \lambda^4\\
0 & \lambda^4 & \lambda^2 \\
\lambda^4 & \lambda^2 & 1\\
\end{array}
\right) \,,
\end{equation}
while the equation on the right leads to two possible textures:
\begin{equation}
M_d^{D(i,I,II)} \sim \left(
\begin{array}{ccc}
0 & \lambda^3 & 0\\
\lambda^3 & \lambda^2 & \lambda \\
0 & \lambda & 1\\
\end{array}
\right)  \qquad
M_u^{D(IV,V)} \sim \left(
\begin{array}{ccc}
0 & \lambda^3 & 0\\
\lambda^3 & \lambda^2 & \lambda \\
0 & \lambda & 1\\
\end{array}
\right) \,.
\end{equation}
Finally the more general case of eq.(18) implies two textures
which are identical to the previous ones,
$M_e^{D(i,I,II)} = M_d^{D(i,I,II)}$ and
$M_e^{D(IV,V)} = M_d^{D(IV,V)}$ plus an extra case:
\begin{equation}
M_e^{D(\star)}\sim \left(
\begin{array}{ccc}
0 & \lambda^3 & 0\\
\lambda^3 & \lambda & 0\\
0 & 0 & 1\\
\end{array}
\right) \,.
\end{equation}

\subsection{The Majorana neutrino mass texture}
On the other hand, Majorana mass terms
$M_R N_i^c N_j$ need not be
generated the same way the other mass terms have been generated so
far. In compactified string models, due
to the absence of large Higgs representations, righthanded
neutrinos donot get tree-level masses, so all entries
in $M_R$ are due to nonrenormalizable operators,
and nothing is a priori known concerning the particular
texture of $M_R$ or the existence of a possible
hierarchy
in this sector. The only constraints
come from the requirement that the seesaw-suppressed masses of
the ordinary neutrinos should be below the experimental upper limits.
For this, $M_R$ has to be a nonsingular matrix and its scale
should be well above the electroweak scale, or else one has to consider the
case of unstable
neutrinos and the related problems from astrophysical and cosmological bounds
[...].
Therefore in addition to the operators that generated the textures
of the up and down quarks and the charged leptons one will need at least an
extra piece to set the Majorana mass scale. Common examples are operators
containing the heavy Higgses
which have been used for generating the texture in eq.(9),
namely $N_i^c H H N_j$,
whose scale is of ${\cal O}(M_G^2/M_S) \simeq 10^{14}$ GeV multiplied for some
orbifold
suppression factor ${\cal C} \simeq 1-10^{-3}$:
\begin{equation}
R = {\cal C}\, {<H> <H>\over M_S}\, \simeq 10^{11} - 10^{14} {\rm GeV} \,.
\end{equation}
In this case $M_R^{ij}\not= 0$ implies
$\alpha_i + \alpha_j = 4\beta$.

Starting with the 4-zero {\it Ansatz} of eq.(8)
for which only the charges $\alpha_1$ and $\alpha_2$ have been specified one
can fix the charges of the light and heavy
Higgs bosons relative to each other such that there is one entry which is
different from zero. In order to generate
more non-zero entries and thus a nonsingular $M_R$ there are two alternatives
paths.
Either some of the
Majorana entries must be generated perturbatively in a similar way as the
entries in the other mass matrices (this implies a hierarchy
of righthanded neutrino scales) [....], or, one should need more
higher-dimension operators:
$N_i^c H_k H_l N_j$, $(k,l = 1,2)$, containing two heavy Higgs multiplets $H_1$
and $H_2$.
Following this second approach one can generate
three alternative Majorana mass textures containing the maximum number of
(four) zero entries [...] \footnote{Notice that with respect
to ref.[...] we have assumed that the entries in
$M_R$ are all equal or of the same order of magnitude $R$.}:
\begin{equation}
M_R^{(b)} \sim \left(
\begin{array}{ccc}
0 & 0 & R\\
0 & R & 0 \\
R & 0 & 0\\
\end{array}
\right)  \qquad
M_R^{(c)} \sim \left(
\begin{array}{ccc}
0 & R & 0\\
R & 0 & 0 \\
R & 0 & R\\
\end{array}
\right)
\qquad
M_R^{(d)} \sim \left(
\begin{array}{ccc}
R & 0 & 0\\
0 & 0 & R \\
0 & R & 0 \\
\end{array}
\right) \,.
\end{equation}
For the same reason, starting from the 5-zero texture solutions of eqs.(12-15)
two extra
heavy Higgs fields -in addition
to the one which led to eq.(9)- are needed in
order to generate the Majorana textures of above.
There may be factors which multiply the non-zero
entries due to the posibility that the two heavy Higgses do not acquire the
same vacuum expectation value, or due to another cause.

\section{The neutrino mass spectrum}
Given the texture structure of the quark and lepton mass matrices
one can calculate the
neutrino masses and the lepton mixing and trace
back the different mass-mixing patterns to the
positioning of texture zeros  in the mass matrices
and thus to the symmetries of the latter.

When $M_R$ is a nonsingular matrix, the seesaw mechanism leads to three
light neutrinos, which are obtained upon diagonalisation of the reduced
$3\times 3$ mass matrix:
\begin{equation}
M_{\nu}^{eff} \simeq  M^{D\dag} M_R^{-1} M^{D}
\,.
\end{equation}
Furthermore if one makes the {\it Ansatz}:
\begin{equation}
M_R \simeq {\bf 1}\times R \quad {\rm or} \quad M_R\simeq M^{D}
\,,
\end{equation}
as this has been common place in the literature, this obviously leads
to the quadratic seesaw relation:
\begin{equation}
m_{\nu_1} : m_{\nu_2} : m_{\nu_3} = ( z^8 : z^4 : 1 ) m_0
\,,
\end{equation}
where the neutrino masses scale either as the up quark masses squared
\begin{equation}
M_D \simeq M_u \quad z=\lambda^2 \quad m_0 = {m_t^2\over R}
\,,
\end{equation}
or they scale as the down quark masses squared
\begin{equation}
M_D \simeq M_d \quad z=\lambda \quad m_0 = {m_b^2\over R}
\,.
\end{equation}
In refs.[...] it was pointed out that the quadratic seesaw spectrum
of eq.(28) is obtained if the Majorana sector has no symmetries on its
own, which means that it has the same symmetries as the quark sector
or it has no symmetries at all.

In order to study this in more detail let us write
the up and down quark textures of eqs.(8,12-15) in a common form,
by means of a set of parameters $\alpha$, $\beta$, $\gamma$, $\delta$ and
${\tilde \alpha}$, ${\tilde \beta}$, ${\tilde\gamma}$, which for
simplicity can be one or zero:
\begin{equation}
M_u = \left(
\begin{array}{ccc}
0 & \alpha \lambda^6 & \delta \lambda^4 \\
\alpha \lambda^6 & \beta \lambda^4 & \gamma \lambda^2 \\
\delta \lambda^4 & \gamma \lambda^2 & 1
\end{array}
\right) \qquad
M_d = \left(
\begin{array}{rcl}
0 & {\tilde \alpha}\lambda^4 & 0 \\
{\tilde \alpha}\lambda^4 & {\tilde \beta}\lambda^2 & {\tilde\gamma}\lambda \\
0 & {\tilde \gamma}\lambda & 1
\end{array}
\right) \,.
\end{equation}
Setting for example $\delta=0$ or $\alpha=0$
one obtains the {\it Ansatz} {\it \`a la Fritzsch} or {\it \`a la Giudice}.
Then the effective light-neutrino mass matrix
assumes a universal form in terms of the scale $z$
which characterizes the Dirac neutrino texture $M^D$,
and as function of the minors
of the matrix $M_R$ $r_{i = 1, ... 6}$ which
are obtained by omitting the row and column containing the corresponding $R_i$
entry, {\it e.g.},
$r_3 = R_1 R_2 - R_4^2$:
\begin{equation}
{M_{\nu}^{eff}\over m_{\nu_3}^2 /\Delta} =  r_3  \left(
\begin{array}{ccc}
\delta^2 z^4 & \gamma \delta z^3 & \delta z^2 \\
\gamma \delta z^3 & \gamma^2 z^2 & \gamma z \\
\delta z^2 & \gamma z & 1 \\
\end{array}
\right) \qquad
\end{equation}
\begin{equation}
\, + \,
r_6 z  \left(
\begin{array}{ccc}
 & (\beta \delta + \alpha\gamma) z^3  & (\alpha+ \gamma \delta) z^2 \\
(\beta \delta + \alpha\gamma) z^3 & 2 \beta\gamma z^2 & (\gamma^2+\beta) z \\
 (\alpha+ \gamma \delta) z^2 & (\gamma^2+\beta) z & 2\gamma  \\
\end{array}
\right)
\end{equation}
\begin{equation}
\, + \,
r_2 z^2  \left(
\begin{array}{ccc}
\alpha^2 z^4 & \alpha\beta z^3 &
\alpha\gamma z^2 \\
\alpha\beta z^3 & \beta^2 z^2 & \beta\gamma z \\
\alpha\gamma z^2 & \beta\gamma z & \gamma^2 \\
\end{array}
\right)
\end{equation}
\begin{equation}
\qquad \, + \,
r_5 z^2  \left(
\begin{array}{ccc}
 &  & \delta^2 z^2 \\
 & 2\alpha\gamma z^2 & (\alpha+ \gamma \delta) z \\
\delta^2 z^2 & (\alpha+ \gamma \delta) z & 2 \delta \\
\end{array}
\right)
\end{equation}
\begin{equation}
\, + \,
r_4 z^3  \left(
\begin{array}{ccc}
 & \alpha^2 z^3 &  \\
\alpha^2 z^3 & 2\alpha\beta z^2 & (\alpha\gamma + \beta\delta) z \\
 & (\alpha\gamma + \beta\delta) z & 2\gamma\delta \\
\end{array}
\right)
\end{equation}
\begin{equation}
\qquad \, + \,
r_1 z^4  \left(
\begin{array}{ccc}
 &  &  \\
 & \alpha^2 z^2 & \\
 &  &  \delta^2 \\
\end{array}
\right)
\end{equation}

If $r_3\not= 0$ and the entries
of the Majorana mass matrix are all of the same order of magnitude the texture
of $M_{\nu}^{eff}$ is, as in the quark sector, either of the {\it Fritzsch}
type with nearest neighbour mixing or of the {\it Giudice} type leading to the
spectrum of eq.(12). In contrast when $r_3 = 0$ as a result of a symmetry
between $N_1$ and $N_2$ the neutrino spectrum can be drastically distorted.
For example for the quark textures of solution .... a texture zero appears
in the (3,3) entry of $M_{\nu}^{eff}$ leading to .....
Notice that in general `anomalous' neutrino mass-mixing patterns can be
obtained only for the five-zero texture solutions of Table 1. The four-zero
solutions require very particular conditions of strong hierarchy in $M_R$
to give mass degenerate neutrinos.

\end{document}